\documentclass[12pt]{article}

\usepackage{epsfig,amsmath,amssymb,latexsym}

\providecommand{\beqa}{\begin{eqnarray}}
\providecommand{\eeqa}{\end{eqnarray}}

\numberwithin{equation}{section}

\def\beq{\begin{eqnarray}}
\def\eeq{\end{eqnarray}}

\def\lsim{\mathrel{\rlap{\lower3pt\hbox{\hskip0pt$\sim$}}
    \raise1pt\hbox{$<$}}}         
\def\gsim{\mathrel{\rlap{\lower4pt\hbox{\hskip1pt$\sim$}}
    \raise1pt\hbox{$>$}}}         

\begin{document}

\thispagestyle{empty}
\rightline{CERN-PH-TH/2008-003, ~LMU-ASC 01/08, ~MPP-2008-6}

\begin{center}

{\bf \LARGE Power of Black Hole Physics: }\\
\vspace{.4truecm}
 {\bf \LARGE Seeing through the Vacuum Landscape}\\

\vspace{1.3truecm}

Gia Dvali$^{a,b}$\footnote{georgi.dvali@cern.ch, gd23@nyu.edu} and Dieter L\"ust$^{c,d,}$\footnote{dieter.luest@lmu.de, luest@mppmu.mpg.de}

\vspace{.6truecm}

{\em $^a$CERN\\
Theory Department\\
1211 Geneva 23, Switzerland}

\vspace{.2truecm}

{\em $^b$Center for Cosmology and Particle Physics\\
Department of Physics, New York University\\
4 Washington Place, New York, NY 10003, USA}

\vspace{.2truecm}

{\em $^c$Arnold Sommerfeld Center for Theoretical Physics\\
Department f\"ur Physik, Ludwig-Maximilians-Universit\"at M\"unchen\\
Theresienstr.~37, 80333 M\"unchen, Germany}

\vspace{.2truecm}

{\em $^d$Max-Planck-Institut f\"ur Physik\\
F\"ohringer Ring 6, 80805 M\"unchen, Germany}

\end{center}

\vspace{1.0truecm}

\begin{abstract}
In this paper we generalize the black hole bound of \cite{dvali}  to de Sitter 
spaces, and apply it to various vacua in the landscape, with a special emphasis  on slow-roll
inflationary vacua.  Non-trivial constraints on the lifetime  and the Hubble expansion rate emerge.  For example, the general tendency is, that for the fixed number and  the increasing mass of the species, vacua must become  more curved and more unstable, either classically or quantum mechanically.  We also discuss the constraints on the lifetime of vacua in the landscape, due to decay into the neighboring states.  
\end{abstract}

\newpage
\pagenumbering{arabic}

\section{Introduction}

It is usually assumed that from the knowledge of low-energy perturbative physics (e.g., such as, the particle spectrum, and their couplings) in our vacuum, one cannot draw any conclusion about the physics in other vacua  on the landscape, without knowing  the non-perturbative structure of underlying high scale theory.  This belief  is based on the intuition, that different vacua correspond to different non-perturbative solutions of the high energy theory, largely separated by the expectation values 
 of the classical order parameters (e.g., vacuum expectation values (VEVs) of the scalar fields),  
 whereas low energy perturbative  physics only accounts for small fluctuations about this solutions. 
As a result, even in the neighboring vacua, physics may be arbitrarily different and unpredictable for a low energy observer in our vacuum.    
 We  wish to  show that  black hole (BH) physics can provide a powerful guideline for overcoming this obstacle. 
Among, the expected enormity  of the vacuum landscape, there is a large subset that shares common gravitational physics.  In these vacua, the classical black hole physics is also common and imposes 
the same consistency constraints on perturbative particle physics.  

 In particular,  by incorporating the consistency bounds, that BH physics imposes on number and masses of particle species \cite{dvali},  we can derive non-trivial constraints not only on our vacuum, but on any quasi-stationary state, which can be obtained by a continuous deformation  of it.   Under {\it continuous deformation}, we mean a change of expectation values that preserves invariant characteristics of the vacuum (such as,   the number of species, 
their chirality, and possibly other topological characteristics).   
 In a certain well-defined sense, to be made precise below,  BH physics allows us to ``see" through the landscape.  
 In this part of the discussion, the key tool in our consideration will be a BH constraint on number of particle species and their masses.  This bound can be derived from the flat space thought experiment, with BH formation and evaporation.   In this experiment, an observer  forms a classical BH and later  
detects its evaporation products.  In each case, when the lifetime of a BH is less than the lifetime of the species, a powerful bound follows.  For example, in the simplest case  the number of stable species of mass $M$ cannot exceed 
\begin{equation}
\label{nmax}
N_{max} \, \equiv \, {M_P^2 \over M^2}\, .
\end{equation}
 This consistency constraint must be satisfied in every vacuum of the theory.   This fact automatically  limits the number of possible deformations of our vacuum, which 
from perturbative physics alone  one would never guess.  For example,  in our vacuum, {\it a priory}, we may  have a very large  number of massless  species coupled to a modulus $\phi$.   Naively,  nothing forbids existence of another vacuum,  obtained by giving an arbitrary VEV to the modulus $\phi$.  However, since such a deformation of the vacuum gives masses to the species coupled to $\phi$, only 
deformations permitted by the BH bound are possible. Thus, BH physics, automatically constraints 
physics in such vacua.  The vacua in question does not  have to be degenerate with ours, or even be stationary.  Below we shall generalize BH bound for such vacua.   Primary target of this study will be 
the de Sitter and quasi de Sitter vacua, that may be connected to ours by a continuous deformation of some  scalar VEVs.  The phenomenological importance of this study is obvious. 
Existence of such vacua is suggested by the strong cosmological evidence that our Universe  underwent a period of inflation, which is responsible for solving the flatness and the horizon problems, and creating the spectrum of density perturbations.  Knowing that we, most likely,  rolled down from another vacuum, we wish  to understand constraints on such states by using BH physics, and whatever knowledge of perturbative physics we have in our present vacuum. 
The bounds from BH physics, which we discuss in this paper, 
set  powerful criteria about
 what is the class of effective string actions, which can
be consistently coupled to quantum gravity, and eventually capture string physics, which might
have been lost in the effective action approach. Those effective field theories
or vacua which cannot fulfill this criterion are called swampland \cite{Vafa:2005ui} (see also
\cite{ArkaniHamed:2006dz}).

 Our  generalization of the BH bound of \cite{dvali} to the de Sitter and quasi de Sitter vacua  relies on  certain relations between the Schwarzschild radius  and the lifetime of a ``test"  BH, and  the Hubble radius and the lifetime of the corresponding  (quasi) de Sitter vacuum respectively.  
  Shortly, for a given number and masses of species, there is an upper limit on the lifetime and the Hubble size of the vacuum, or else the BH bound (\ref{nmax}) must be satisfied.    In the other words, a given vacuum can only invalidate this  BH bound on species, by becoming more curved and/or  shorter lived. For the slow-roll inflationary vacua,  this implies constraints on the slow-roll parameters, and subsequently, on the allowed number of the inflationary e-foldings.  

 For the classically-stable vacua, the story is a bit more subtle. Naively, since such vacua are exponentially long lived, their lifetime should exceed the lifetime of any sensible BH that can fit within 
 their de Sitter horizon.   Hence, such vacua should automatically fall within the validity of our arguments, and the only resulting constraint should be on their curvature scale.  However, in the light of enormity of the string landscape, the tunneling rate can be enhanced by number of the neighboring vacua, to which they can decay via 
 quantum-mechanical tunneling.  Intuitively it is clear,  that at least in weakly-coupled theories, 
 the exponential longevity of meta-stable vacua  should be maintained, or else  the vacuum in question can no longer be treated as a well defined quasi-classical state.   However,  for our purposes in many cases the issue may be  quantitative and  we therefore perform a brief investigation of this question in the first part of the paper.

Thus, in the first part of the paper we consider constraints on the lifetime of vacua. 
The purpose of this study is twofold. First, as explained above,  we wish to make sure that 
enormity of the possible decay channels does not interfere with generalization of the BH bound to 
such meta-stable vacua.   
Secondly,  there is a phenomenological byproduct.  Demanding 
 that our vacuum with cosmological constant
$\Lambda$ has a long enough life-time, one gets some
constraints on $\Lambda$ which are related to the number $N_{vac}$ of vacua.
As a result we will see that an upper bound on $\Lambda$ can be obtained, which 
decreases with the number $N_{vac}$ of vacua (or conversely an upper limit on $N_{vac}$ arises,
which decreases with $\Lambda$). 
So, if there exist too many vacua, the cosmological constant must be smaller than its 
observed value today.
As expected, for our present vacuum, these bounds are rather mild, and become
only phenomenologically interesting for very large numbers of $N_{vac}$. 

In string theory, vacuum decay processes are related to domain wall configurations.
These domain walls can be thought to be built by intersecting (D)-branes of the
underlying superstring
theory.  We will discuss some aspects of vacuum decays in the string landscape, in particular
also the possibility of tunneling from de Sitter or Minkowski vacua to anti-de Sitter vacua, which
typically arise in flux compactifications. The corresponding domain wall solutions are given by branes
that precisely act as sources for the background fluxes.
In this way we can derive some constraints on the flux quantum numbers, by requiring Minkowski or
de Sitter vacua with long enough life-time.

In section 3 we discuss bounds on the landscape of effective field theories from BH decays.
 As it was discussed in \cite{dvali} and also in \cite{qg} these bounds provide
a possible  explanation 
of various hierarchies observed in nature. Namely
in \cite{dvali} some perturbative and non-perturbative arguments were given that in
a quantum field theory with $N$ species of particles of mass $M$,  there
is for large $N$  an inevitable hierarchy between the  Planck mass $M_P$ and $M$:
\begin{equation}
\label{boundN}
M_P^2\geq N \, M^2\,.
\end{equation}
In particular, for $N$ of order $N\sim 10^{32}$, the bound (\ref{boundN}) 
explains the hierarchy between $M_P$ and the $N$ species roughly at the TeV-scale.
E.g. this large number of particles is realized in scenarios with large number of extra
dimensions and $10^{32}$ KK modes at the TeV-scale \cite{Arkani-Hamed:1998rs}.
 Recently \cite{strongcp}, it was argued, that large number of standard-model-like species also leads to the smallness of strong CP parameter. 

We will generalize the BH bounds to the case of non-static de Sitter Universes and to 
quasi de Sitter type time-dependent backgrounds.  This will give us new restrictions on
inflationary scenarios like chaotic inflation or D-brane inflation. 
We also consider bounds from BH on landscape models with softly
broken supersymmetry in static Universe.  We shall see, that the large number and the mass of the species tends to make the vacuum more highly curved and shorter lived. 

  Finally, we wish to stress an important point concerning the possible relevance of the lowered cutoff of the theory for the generalized BH bound. As it was shown in \cite{qg},  with increasing number of species, not only their masses, but also the gravitational cutoff of the theory gets lowered and is bounded from above by $M_P/\sqrt{N}$.  In particular, this conclusion agrees with the perturbative argument \cite{dvaligabadadze, veneziano} about the one-loop renormalization of the Planck mass by $N$ species.  In our constraint of the de Sitter vacua, the central role is played by the bound (\ref{boundN}), which has to be satisfied by all the relevant (long enough lived species that can fit within the appropriate  BH horizon (see below)) species, irrespectively whether they are above of below the cutoff.   This fact is important for our applications to the string landscape, as  
it allows us to constrain vacua in which the masses of the species are way above the string scale, although the latter is the cutoff of the theory.  For instance, such are the brane-inflationary vacua in which the heavy species correspond to the  lowest excitations of the stretched strings.  Such states, although 
they are heavier than the string scale,  fit within the sub-horizon BH, and therefore fall within the validity of the bound \cite{dvali}.

\section{Constraints on the life-time of vacua}

In this section we provide a discussion about possible restrictions on
the maximal number of vacua resp. on the cosmological constant, considering 
two kinds of decay processes.
First we will consider transition between vacua with positive cosmological constant
via the creation of expanding bubble. These bubbles are created by Coleman/De Luccia
gravitational effects \cite{Coleman:1980aw} resp. by Hawking/Moss instantons \cite{Hawking:1982my}. 
Then we will discuss the string landscape which typically also contains a large number
of AdS vacua with negative cosmological constant. We consider domain wall solutions
(membranes in four dimensions) which in analogy to the Coleman/De Luccia instantons
can create bubbles of contracting universes, and hence can be responsible
for the decay of a Minkowski or de Sitter  vacuum into a vacuum with negative
cosmological constant.


\subsection{De Sitter vacuum decay by quantum tunneling}

\subsubsection{Creation of  a single bubble}

It is known from the work of Coleman  and De Luccia \cite{Coleman:1980aw}
that a de Sitter universe with cosmological constant $\Lambda \equiv V_0$
can decay into another vacuum de Sitter vacuum or into a
Minkowski vacuum which are separated from each other by a 
potential barrier of height $V_1$. 
In Euclidean quantum gravity, the de Sitter entropy of an expanding universe
with vacuum energy $V_0$
is determined by the value of the classical action:
\begin{equation}S_0={24\pi^2\over V_0}\, .
\end{equation}
Using this value for the Euclidean action, one can compute
the 1-instanton decay rate of this De Sitter universe into another vacuum via quantum tunneling
in a semiclassical approximation\footnote{This amplitude was also used in \cite{Kachru:2003aw}
to show that a {\sl single} KKLT vacuum in type IIB superstrings has a life-time
longer than the age of the universe.}:
\begin{equation}
\Gamma_{(1)}\simeq
M_P\exp\biggl( -{24\pi^2M_P^4\over V_0}+{24\pi^2M_P^4\over V_1}\biggr)\, .
\end{equation}
Assuming that the height of the barrier is much bigger that $\Lambda$, $V_1>>V_0$, one
simply obtains for the life time $\tau$ resp. the decay rate $\Gamma$:
\begin{equation}
\tau^{-1}\sim\Gamma_{(1)}\simeq
M_P\exp\biggl( -{24\pi^2M_P^4\over \Lambda}\biggr)\, .
\end{equation}

Note that it is also possible that a quantum jump from $V_0$ to the top of the barrier $V_1$, which
is followed by an decay to another de Sitter vacuum with cosmological constant $V_2$, where
$V_2>V_0$. This was discussed by Hawking and Moss and is also closely related
to thermodynamic fluctuations due to the de Sitter entropy of the vacuum with cosmological
constant $V_0$. Specifically, 
the decay rate of our vacuum by creation of a single new bubble is given by:
\begin{equation}
\tau^{-1}\sim\tilde\Gamma_{(1)}\simeq M_Pe^{-{E\over T_H}}\, ,
\end{equation}
where $E$ is the energy necessary to thermally create the new bubble, and
$T_H$ is the Hawking temperature of our de Sitter universe:
\begin{equation}
T_H\simeq{\sqrt{V_0}\over M_P}={\sqrt\Lambda\over M_P}\, .
\end{equation}

\subsubsection{Decay into $N_{vac}$ vacua}

Now we want to consider a much bigger landscape of $N_{vac}$ different vacua, into which our universe can decay
via quantum tunneling. 
First, we consider the case, where all different vacua can be reached by a single tunneling process.
Adding up all these 1-instanton decays into the $N_{vac}$
different vacua one simply obtains the following decay amplitude:
\begin{equation}
\Gamma_{(N_{vac})}\simeq
N_{vac}M_P\exp\biggl( -{24\pi^2M_P^4\over \Lambda}\biggr)\, .
\end{equation}
Now requiring that for our universe this decay amplitude is suppressed
such that our universe has a long enough life-time, i.e.
\begin{equation}
\label{bound2b}
\Gamma_{(N_{vac})}<H={\sqrt{\Lambda}\over M_P}\, ,
\end{equation}
we derive the following bound on $\Lambda$:
\begin{equation}\label{bound2d}
\Lambda<{24\pi^2M_P^4\over \ln N_{vac}}\, .
\end{equation}
E.g. for $N_{vac}=10^{500}$ one gets ${\Lambda/(24\pi^2M_P^4)}<8.7\times 10^{-4}$, whereas as for 
$N_{vac}=10^{1500}$ one gets ${\Lambda/(24\pi^2M_P^4)}<2.9\times 10^{-4}$.
Again, the obtained bounds on $\Lambda$ are not very exciting, unless the landscape is extremely
huge.


However we should  consider not only the 1-instanton process, but also all k-instanton processes,
which describe the process that we can reach a certain bubble via the subsequent decay over $k$
different bubbles.
%
In a kind of instanton dilute gas approximation one gets for each step a suppression factor of
$e^{-S_0}$, and hence the decay amplitude for reaching one specific vacuum via  $k$ tunneling processes (k-instanton process) becomes
\begin{equation}
\Gamma^k\simeq
M_P\exp\biggl( -{24\pi^2kM_P^4\over \Lambda}\biggr)\, .
\end{equation}
In order to obtain the full decay amplitude into $N_{vac}$ different vacua we sum over all
possible k-instanton processes, i.e. taking into account all possible ways
decay processes. 
Then
we finally obtain:
\begin{eqnarray}\label{total}
\Gamma^{\rm total}_{(N_{vac})}&\simeq &
M_P\biggl(\sum_{k=0}^{N_{vac}}{N_{vac}(N_{vac}-1)\dots (N_{vac}-k)\over k!}
\exp\biggl( -{24\pi^2kM_P^4\over \Lambda}\biggr)-1\biggr)
\nonumber\\
&=&
M_P\biggl(1+  e^{ -{24\pi^2M_P^4\over \Lambda}}\biggr)^{N_{vac} }-M_P
\, .
\end{eqnarray}
Requiring that that our universe with cosmological constant $\Lambda$ is stable enough,
\begin{equation}
\label{bound2}
\Gamma^{\rm total}_{(N_{vac})}<H={\sqrt{\Lambda}\over M_P}\, .
\end{equation}
we obtain again an upper bound on $\Lambda$ which now reads
\begin{equation}
\Lambda<{24\pi^2M_P^4\over \ln (N_{vac}/\ln 2)}\, .
\end{equation}
This essentially agrees with the bound eq.(\ref{bound2d}) obtained before.

\subsection{Decay of vacua in the landscape of string flux compactications}

\subsubsection{Vacuum decay for fixed background  fluxes}

As it is well know string compactifications lead to a huge number of lower
dimensional ground states \cite{Lerche:1986cx,Bousso:2000xa,Douglas:2003um}. In particular the number $N_{vac}$ of discrete vacua 
in the context of flux compactifications of type II orientifolds was
estimated to be of order of $N_{vac}\sim 10^{500}$. 
Therefore a statistical analysis of flux vacua was suggested in \cite{Douglas:2003um,Denef:2007pq}.
Wrapping in addition D-branes
around cycles of the underlying (Calabi-Yau) spaces in order
to derive the Standard Model of particle physics increases this number even further.
Therefore intersecting brane models and the likelihood to derive the Standard
Model were also investigated in a statistical manner \cite{Blumenhagen:2004xx,Dijkstra:2004cc,Gmeiner:2005vz,Douglas:2006xy,Gmeiner:2007we,Gmeiner:2007zz}.
Here we want to discuss some constraints on the landscape
of type II compactifications with p-form fluxes and also possible non-perturbative effects like
gaugino condensation and Euclidean instantons, as it was proposed first in the KKLT scenario
\cite{Kachru:2003aw}.

We will discuss flux compactifications in the context of the effective supergravity action.
In a general ${\cal N}=1$ supergravity, the scalar potential $V$
 is a function of chiral superfields $\phi_i$ and
 takes the standard form
\begin{equation}\label{scapot}
V=\mathrm{e}^K\left(|D_iW|^2-3|W|^2\right)+|D_a|^2\, ,
\end{equation}
where $D_a$ are the D-terms, and the F-terms are defined as
\begin{equation}\label{fterm}
F_i=\mathrm{e}^{K/2}D_iW=\mathrm{e}^{K/2}\left(
\partial_{\phi_i}W+W\partial_{\phi_i}K\right)
\end{equation}
with $W$ being the superpotential and $K$ the K\"ahler potential.

Our aim is to find local minima of $V$. We must
therefore impose
\begin{equation}\label{ftermmin}
{\partial V/\partial \phi_i}|_{\phi_{\mathrm{min}}}=0 \ \forall  i\, .
\end{equation}
Supersymmetric minima are obtained if all $F_i|_{\phi_{\mathrm{min}}}=D_a|_{\phi_{\mathrm{min}}}=0$.

Let us  neglect the possible contribution of D-terms to the scalar potential.
In this case $V$ is fully specified by the K\"ahler potential $K$ and the superpotential $W$ in 
eq.(\ref{fterm}).
The generic form of the superpotential in type II orientifold compactifications is of the form
\begin{equation}
W=W_{\rm flux}(\phi)+W_{n.p.}(\phi)\, .
\end{equation}

On a generic  Calabi-Yau space the total number of (type IIB) flux vacua is
estimated by the following equation \cite{Bousso:2000xa,Ashok:2003gk,Denef:2004ze}:
\begin{equation}\label{number}
N_{SUSY}\simeq { L^{2h^{2,1}+2}\over (2h^{2,1}+2)!}\, 
\end{equation}
Here the Hodge number $h^{2,1}$ counts the number of complex structure moduli,
and $L$ is the orientifold charge of the system. Typical numbers for $h^{2,1}$ and $L$
indeed lead to a huge number of supersymmetric flux vacua.
This number counts all different 3-form flux combinations that lead to a
supersymmetric ground state satisfying that lead to a solution of the supersymmetry equations
with respect to $W_{\rm flux}$:
\begin{equation}\label{SUSYflux}
D_{\phi}W_{\rm flux}=0\, 
\end{equation}
Including the non-perturabtive superpotential 
$W_{n.p.}$ and also looking for non-supersym\-metric
local, i.e. metastable vacua will not change this number by a considerable amount, i.e.
the total number $N_{vac}$ of local string vacua is comparable to $N_{SUSY}$: $N_{vac}\simeq N_{SUSY}$.
In particular eq.(\ref{number}) means that the huge number of flux vacua originates
from the big number of possibilities of choosing different flux vectors though the homology
3-cycles of the CY space.

Let us first consider transitions between vacua with fixed values for the fluxes,
i.e. all vacua have the same flux quantum numbers.
These transitions are due to gravitational non-perturbative effects, like e.g. the Coleman/De Luccia
instantons in case of positive cosmological constants, as describes above.
However, since the fluxes are quantized and hence take discrete values, there exist only
a few transitions that are possible.
Indeed, on a given moduli space of
type IIB complex structure moduli $\phi^{2,1}$, K\"ahler
moduli $\phi^{1,1}$  and including the dilaton $\tau$, transitions
between different vacua are only possible for fixed background fluxes,
and also for  fixed non-perturbative effects. 
I.e. fixing the flux parameters and also the non-perturbative superpotential,
the corresponding scalar potential has only a relatively small number of (local) minima,
denoted by $N^*$, and in general one has that
$N^*<<N_{vac}$.
This
fact largely restricts the possible vacuum decay processes within the string
flux landscape for fixed fluxes. Varying the moduli fields, only a small subset of vacua can be reached
by vacuum tunneling and decay processes, along the lines described in section two. 
E.g. in type IIB flux compactifications
supersymmetric solutions are characterized by imaginary self-dual fluxes
$G_3$ \cite{Giddings:2001yu}, whereas nonsymmetric local minima of $V$
allow for more general flux choices.
The extremality conditions  comprise $h^{2,1}+1$ conditions for $h^{2,1}+1$ complex variables.
Therefore one expects that the degeneracy in the moduli space is in general
totally lifted, and one obtains a discrete set of solutions for the moduli fields.
Their number $N^*$ depends on the prepotential $F(U)$ of the underlying Calabi-Yau manifold.
As one can show the number $N^*$ of solutions of eqs. ({\ref{SUSYflux})  is essentially of order one.
E.g. consider a GVW/TV superpotential \cite{Gukov:1999ya,Taylor:1999ii} of the form \cite{Curio:2000sc}
\begin{equation}
W_{\mathbb{IIB}}=(p+iqSU_1)(l_2-il_1U_2+in_1U_3-n_2U_2U_3)\, .
\end{equation}
$p,q,l_1,l_2,n_2,n_2$ parametrize the flux quantum numbers that are constrined by the
tadpole condition.
For fixed flux quantum numbers there is a unique solution of the supersymmetry condition
with zero vacuum energy:
\begin{equation}
SU_1=-{p \over q}\, , \quad U_2=\sqrt{l_1l_2\over n_1n_2} 
\, ,\quad U_3=\sqrt{l_2n_1\over l_1n_2} \, .
\end{equation}

How many other local (non-supersymetric) minima of $V$ may exist besides the supersymmetric
Minkowski or $AdS_4$ groundstates? The answer to this question in general depends
on the details of the non-perturbative part of the superpotential and also on the up-lift
procedure, e.g. by additional D-terms or non-supersymmetric contributions
to the potential.  In general, we expect that the total number $N_{vac}$  of vacua  of different possible flux choices
is by far larger that the number $N^*$ of local minima of the scalar potential
with fixed fluxes.
This can be seen as follows: In KKLT \cite{Kachru:2003aw} the modification of the IIB flux superpotential
by non-perturbative D-instantons or by gaugino condensation is of the following form:
\begin{equation}
W=W_0(U)+A(U)e^{-aT}\, .
\end{equation}
The fluxes entirely enter in $W_0$, which can be treated in some approximation as
a constant contribution to the superpotential. Each different choice for the fluxes leads
to some specific $W_0$. However for given fluxes, i.e. given $W_0$, the number of
local minima of the scalar potential is low. If we vary all moduli parameters plus
the dilaton field, we are moving in a  moduli space ${\cal M}$ of (complex) dimension
$dim({\cal M})=(h^{1,1}+h^{2,1}+1)$. 
On general grounds, we expect that the number $N^*$ of solutions
of eq.(\ref{ftermmin}) is at most of the order $dim({\cal M})$. This is obviously smaller that the number 
$N_{SUSY}$ given
eq.(\ref{number}). 
Finally uplifting the potential by a small amount, in oder to obtain a vacuum with small
cosmological constant $V_0$, will not drastically change the number $N^*$ of metastable
vacua.

\subsubsection{Vacuum decay due to stringy domain walls}

In order to get transitions between vacua with different flux quantum numbers,
one needs non-perturbative, gravitational configurations which are coupled
to the flux background fields, and which interpolate between different flux
vacua. These are given in term of BPS or nearly BPS domain walls (membranes) 
(for earlier work see e.g. \cite{Cvetic:1991vp,Cvetic:1996vr})
in four-dimensional
space time that are coupled to the scalar moduli fields.
The profile of the domain wall is such that it separates spatial regions with different flux quantum
numbers from each other. For the case that the domain wall is interpolating between two 
supersymmetric vacua, the interpolating solutions is describing a BPS domain wall.
Of course, eventually we are interested in the decay of a non-supersymmetric flux
vacuum with positive cosmological constant (our vacuum) and broken space-time supersymmetry
into another (supersymmetric) flux vacuum, which cam have either positive, zero or also
negative cosmological constant ($AdS_4$) vacuum. The formation of an $AdS_4$ domain wall
is particularly interesting, since $AdS_4$ are very common in the string landscape.
In this case our universe would be decaying into a contracting space, which at first sight seems
to be problematic. Nevertheless the corresponding transition amplitude 
from $dS_4$ to $AdS_4$ is expected to be non-vanishing,
as it was  discussed in \cite{Ceresole:2006iq}.

To demonstrate a
vacuum transition
between string vacua with different fluxes, we discuss as a simple  example we  type IIA, $AdS_4$ flux vacua with all moduli fixed at finite values. The corresponding domain walls were recently constructed 
in \cite{Kounnas:2007dd}, and they are microscopically composed of  intersecting D-branes, NS 5-branes and possibly also
by socalled Kaluza-Klein monopoles.
Specifically, consider a flux superpotential of the form 
\cite{Derendinger:2004jn,Villadoro:2005cu, DeWolfe:2005uu, Camara:2005dc,Kounnas:2007dd}
\begin{equation}\label{IIAfull}
W_{\mathrm{IIA}}=W_H+W_F \, .
\end{equation}
The first term is due to the Neveu--Schwarz 3-form fluxes and
depends on the dilaton $S$ and the type  IIA complex-structure
moduli $U_m$ ($m=1,\dots ,\tilde h^{2,1}$):
\begin{equation}W_H(S,U)=\int_Y\Omega_c\wedge H_3=i\tilde a_0S+i\tilde c_mU_m\, ,
\end{equation}
where in type IIA the 3-form $\Omega_c$ is defined by
$\Omega_c=C_3+i\mathrm{Re}(C\Omega)$. Second, we have the
contribution from Ramond 0-, 2-, 4-, 6-form fluxes:
\begin{eqnarray}
W_F(T) &=&\int_Y \mathrm{e}^{J_c}\wedge F^{\mathrm{R}}\nonumber
\\ &=&\tilde m_0\frac{1}{6}\int_Y\left(J_c\wedge J_c\wedge
J_c\right) +\frac{1}{2}\int_Y\left(F_2^{\mathrm{R}}\wedge
J_c\wedge J_c\right)+
\int_YF_4^{\mathrm{R}}\wedge J_c+\int_YF_6^{\mathrm{R}}\nonumber \\
&=&i\tilde m_0F_0(T)-\tilde m_iF_i(T)+i\tilde e_iT_i+\tilde
e_0\,\label{IIAF} .
\end{eqnarray}
Here $F(T):= F_0(T)$ is the type IIA prepotential, which depends
on the IIA K\"ahler moduli $T_i$ ($i=1,\dots ,\tilde h^{1,1}$) and
$F_i(T):=\partial F_0/\partial T_i$. We use the notation $J_c$ for
the complexified K\"{a}hler metric $J_c:=B+iJ$. 
Assuming a simple (toroidal) cubic prepotential $F=T_1T_2T_3$, the
superpotential has the generic form:
\begin{eqnarray}
W_{\mathrm{IIA}}&=&W_F+W_H= \tilde m_0\int_Y(J\wedge J\wedge J)+
\int_YF_4^{\mathrm{R}}\wedge J+
\int_Y\Omega_c\wedge H_3\nonumber\\
&=&i\tilde e_iT_i+i\tilde m_0T_1T_2T_3+i\tilde a_0S+i\tilde
c_mU_m\, .\label{IIAU}
\end{eqnarray}
With $K= - \log (S+\bar S) \prod_{i=1}^3(T_i+\bar
T_i)\prod_{i=1}^3(U_i+\bar U_i)$, the equations (\ref{SUSYflux})
admit the following unique solution with all moduli stabilized:
\begin{equation}\label{solutionIIAg}
|\gamma_i|T_i
=\sqrt{\frac{5|\gamma_1\gamma_2\gamma_3|}{3 \tilde m_0^2}}\,
,\quad S=-\frac{2}{3 \tilde m_0\tilde a_0}\gamma_iT_i\, ,\quad
\tilde c_m U_m=-\frac{2}{3 \tilde m_0}\gamma_i T_i\, , \quad \gamma_i=\tilde m_0\tilde e_i\, .
\end{equation}
This solution corresponds  to supersymmetric $\mathrm{AdS}_4$
vacuum with negative cosmological constant:
\begin{equation}
\Lambda_{AdS}=-3e^K|W|^2= -{3^7\sqrt{3\over 5}\over 100}{|\tilde a_0\tilde c_1\tilde c_2\tilde c_3|(|\tilde m_0\tilde e_1\tilde e_2 \tilde e_3|)^{5/2}\over(\tilde e_1\tilde e_2\tilde e_3)^4}M_P^4\, .   \label{lambdaads}
\end{equation}

Now let us consider the corresponding the domain wall solution which interpolates 
between the above $AdS_4$ flux vacuum and flat Minkowski space-time with vanishing
fluxes. As discussed in \cite{Kounnas:2007dd} it is given in terms of intersting D4,- D8- and NS 5-branes.
In addition one also needs orientifold 6-planes (O6) in order to cancel the induces
D6-brane charge from the fluxes. The complete form of the 10-dimensional metric as well as the
profiles of the scalar fields can be found in  \cite{Kounnas:2007dd}.
The
four dimensional part of the metric is such of an interpolating domain wall, where the
intersecting branes are smeared in the direction transveral to the domain wall.
Specifically, this 4-dimensional part of the metric can be written as 
\begin{equation}
{\rm d}s^2=a(r)^2(-{\rm d}t^2+{\rm d}x^2+{\rm d}y^2)+{\rm d}r^2\, .\label{domainmetric}
\end{equation}
For $r\rightarrow 0$ this metric approaches the metric of $AdS_4$, and the scalar fields
are fixed to the values determined by the non-vanishing fluxes, as given in eq.(\ref{solutionIIAg}).
For $r\rightarrow \infty$, the function $a(r)$ becomes a constant, and the eq.(\ref{domainmetric})
become the metric of flat Minkowski space. 

The tension $\sigma$ of the domain wall can be computed by introducing a central function
$Z(r)$ which is defines as
\begin{equation}
Z(r)={a'(r)\over a(r)}\, .
\end{equation}
By comparison with the exact metric of  \cite{Kounnas:2007dd} one obtains
\begin{equation}
Z(r)|_{r=0}=e^{K/2}|W|\, ,\quad \Lambda_{AdS}=-3|Z(r)|^2_{r=0}\, .
\end{equation}
The (membrane) tension $\sigma$ of the domain wall is then given by the following expression:
\begin{equation}
\sigma\simeq (|Z|_{r=\infty}-|Z|_{r=0})\, .
\label{tension}
\end{equation}

Now let us determine the decay amplitude of the Minkowski vacuum with vanishing fluxes
into the $AdS_4$ vacuum with non-vanishing fluxes. 
The decay of the Minkowski vacuum occurs due to the creation of the domain wall, which speews
through space-time until the entire universe is in the new $AdS_4$ vacuum. This is similar
but not completely equal to the creation of a bubble via the Coleman/De Luccia instanton.
In fact in order to be realistic, one
should break supersymmetry and uplift the Minkowski vacuum by a small amount to
obtain a de Sitter vacuum which decays into the $AdS_4$ vacuum. 
Neglecting the problem of supersymmetry breaking and the uplift, the decay amplitude
of the Minkowki (de Sitter)  vacuum is then given by the
following expression:\footnote{The details of the derivation of this equation by computing
the Euclidean action of the domain wall of ref. \cite{Kounnas:2007dd} coupled to the scalar fields
will be given elsewhere \cite{workinprogress}.}
\begin{equation}
\Gamma \simeq M_P\exp\biggl( -{8\pi^2M_P^4C\over \sigma^2}\biggr)=
M_P\exp\biggl( {24\pi^2M_P^4C\over \Lambda_{AdS}}\biggr)\, .
\end{equation}
The constant $C$ depends on the details of the domain wall solution.

As also discussed in \cite{Ceresole:2006iq}, unlike the cases discussed in section two,
the corresponding decay amplitude is independent of the de Sitter cosmological constant $\Lambda
=V_0$, but  only depends on the value of $\Lambda_{AdS}$.
In order to avoid too fast decay of our vacuum, $|\Lambda_{AdS}|$ must not be too large. 
E.g. if $|\Lambda_{AdS}|\simeq m_{3/2}^4$, the life-time of our universe is long enough.
However $AdS_4$ vacua with $|V_1|\sim M_P^4$  create too much decay of our vacuum.
Using the known expression
for $\Lambda_{AdS}$ in
eq.(\ref{lambdaads}),  this constraint can be translated into the following restriction on the flux 
quantum numbers:
\begin{equation}
{3^7\sqrt{3\over 5}\over 100}{|\tilde a_0\tilde c_1\tilde c_2\tilde c_3|(|\tilde m_0\tilde e_1\tilde e_2 \tilde e_3|)^{5/2}\over(\tilde e_1\tilde e_2\tilde e_3)^4}<<1\, .
\end{equation}

\section{Black Hole Proof for de Sitter} 

  Before studying applications of the bound on species (\ref{nmax}) to the vacuum landscape,  we wish to generalize the BH proof of the bound to the de Sitter and quasi de Sitter spaces.  Let $M$ be the mass of the species, and let $H$ be the Hubble parameter in de Sitter. 
We wish to perform a thought experiment  \cite{dvali}, in which number of species is  absorbed by a BH, which then evaporates and releases them back.  
The key point is, that the BH can start emitting the species only after its Hawking temperature
becomes comparable to their mass, and this fact implies (\ref{nmax}).     
 The necessary requirement 
for such an experiment is that the gravitational radius $r_g\equiv M_{BH}/M_P^2$ of the  BH of the interest, must be 
less than the Hubble radius
\begin{equation}
\label{randh}
r_g \, \ll \, H^{-1}.  
\end{equation}
We shall split the rest of the discussion into two parts, by imposing different constraints on the BH lifetime.  This is dictated by the fact that for the validity of such an experiment, not only the size but the lifetime  of the BH also matters.    What is important, is that vacuum itself must be longer lived than the BH.  This implies different constraints on the type of BH that we can use in our analysis for 
the vacua with different  level of time-dependence.

\subsection{Time-Dependent Vacua:  Constraints from Short-Lived Black Holes}

In the first case, let us require that not only the gravitational radius, but also the lifetime of the BH be less 
than the Hubble time.  That is, 
\begin{equation}
\label{lifetime}
\tau_{BH}  \, \ll \, H^{-1}. 
\end{equation}
 Notice, that the lifetime of a black hole depends on the number of species into which it can evaporate, 
 and which in our case may be very large.   For small black holes,  $r_g \, \ll \, M^{-1}$, this correction can be very important, and must be taken into the account.   On the other hand, for  large BH  $r_g \, \gg \, M^{-1}$, that mostly evaporate into few very light species (such as a graviton or a photon), the correction 
 to the lifetime from $N$ heavy states is unimportant, and for these BH the life-time is approximately given by   
\begin{equation}
\label{lifetime1}
\tau_{BH}  \, \sim \, r_g^3M_P^2. 
\end{equation}
  
In practical applications, the requirement (\ref{lifetime})  will be relevant for the vacua that have a relatively short life-time, e.g., such as the slow-roll inflationary vacua,  which can be regarded as stationary only for several  Hubble times.  

Notice, that since in any sensible (quasi) de Sitter state  $H^{-1}M_P \, \gg \, 1$, the condition (\ref{lifetime}) outomatically implies (\ref{randh}). 
That is, a BH that evaporates in less than a Hubble time, is automatically small enough to fit within the Hubble horizon.  Let us now prepare such a BH, by putting together $n$ particles, all from  different species.  The maximal number of particles that we can add to a BH,  without violating the requirement (\ref{lifetime}) (and automatically (\ref{randh})), is limited by the following consideration.

 In order to fit a particle into a BH, the typical momentum of the particle (that is, its characteristic  inverse localization width) must be higher than $r_g^{-1}$.  Indeed, even if a particle in question  is massless,  in order to throw it into a BH, we have to prepare a localized wave-packet  of the size $\Delta X \lsim r_g$. 
Such a wave-packet will have a characteristic  momentum  $\Delta P \, \gsim \, r_g^{-1}$.  Thus,  throwing a particle of the rest mass $M$ 
into a BH, we automatically increase the mass of the latter  at least by $\Delta M_{BH} \, \simeq \,
 \sqrt{M^2 \, + \, r_g^{-2}}$, and correspondingly, increase its horizon by 
 \begin{equation}
\label{deltar}
 \Delta r_g \, \sim  \,
{ \sqrt{M^2 \, + \, r_g^{-2}} \over M_P^2}.
\end{equation}
 (Notice, that the converse is also true.  When a black hole emits a  particle, due to the thermal nature of Hawking radiation, the typical energy released is $\sim r_g^{-1}$, and decrease in the horizon is (\ref{deltar})).  
   To find the BH mass as a function of number $n$ of the  ``constitutent" particles,  we must summ over all the increments. Approximating the sum by the integral, we get 
 the following expression for  the number of particles necessary for building a BH of a given  mass $M_{BH} $, 
\begin{equation}
\label{nandm}
n(M_{BH}) \, \simeq \, \int_0^{M_{BH}}\, {dm \over \sqrt{M^2 \, + \, M_P^4/m^2}}  \,  
=\, {1\over 2M^2} \left(\sqrt{M^2M_{BH}^2 \, + \, M_P^4} \, - \, \ M_P^2 \right ).     
\end{equation}
The maximum number of particle species, ${\bar n}$,  that can participate in our experiment, is set by the number of particles  that is needed to grow the BH to a critical mass,  $M_C$, with the 
lifetime becoming comparable to $H^{-1}$. 
That is,  ${\bar n} \equiv n(M_C)$, where $M_C$ saturates the bound (\ref{lifetime}). 
At this point, it is usefull to split the discussion into two parts, corresponding to the cases when 
$M \, \gg \, H$, and $M \, \lsim \, H$. 
   
 \subsubsection{Constraint on Heavy Species:  $M \, \gg \, H$}
 
   This case requires a careful analysis, since the black hole lifetime, which is a function of its 
size $r_g$,  undergoes an abrupt transition around the critical size $r_g \, \sim \, M^{-1}$.  The reason is, that the black holes of size $r_g \, \lsim \, M^{-1}$ have Hawking temperature $T_H \, \gsim \, M$, and can radiate all the constituent species. So the lifetime of a subcritical BH  is 
\begin{equation}
\label{smallbh}
\tau_{BH} (r_g \lsim M) \, \sim \, {1 \over n}  {M_{BH}^3 \over M_P^4}\, .  
\end{equation}
Recall that our experiment is designed in such a way, that we are forming a minimal BH out of $n$  particles belonging to {\it different} species.  
So due to conservation of species number, such a minimal  BH can only radiate the $n$  particles belonging to the  input species, and not other energetically available $N \, - \, n $ species.  In order to find number of species needed for  building a BH of size $r_g \, \sim \, M^{-1}$, we just have to take $M_{BH} \, = \, M_P^2/M$ in eq(\ref{nandm}). 
Ignoring the factors of order one, this gives
 \begin{equation}
\label{nsmall}
n(M_P^2/M) \, \sim \, {M^2_P \over   M^2}.     
\end{equation}
Plugging this into the eq(\ref{smallbh}), we get 
\begin{equation}
\label{smalltau}
\tau_{BH} \, \sim  M^{-1}.
\end{equation}  
This result is already indicative. Equation (\ref{nsmall}) is compatible with the flat space bound, which 
is certainly applicable because the lifetime of a BH is $M^{-1} \, \ll \, H^{-1}$.  Moreover, the fact that $N$ cannot exceed $M_p^2/M^2$  can be anticipated from the fact that if it could, we could form
a neutral BH of mass $M^{-1}$, with even less lifetime.  This would indicate that such BHs cannot be treated as well defined states, in agreement with the result of  \cite{qg}. 

 So far,  what we know is, that a BH of size $\sim M^{-1}$ can contain maximum 
 $M_P^2/M^2$ units of the conserved species number. 
 Now,  let us try to build a bigger BH by putting in more 
 species (as before, we keep adding particles from the different species!).   Once the BH grows 
 $r_g \gg M^{-1}$, there is sharp increase in its lifetime, since the emission of states with mass 
 $M$ becomes  exponentially suppressed by the Boltzmann factor.  At this point, the BH can only evaporate 
 into the small number of available massless species (such as the graviton), and the lifetime is given by 
 (\ref{lifetime1}).


Requiring that the resulting BH satisfies the lifetime constraint (\ref{lifetime}),  that is, 
\begin{equation}
\label{big}
r_g \, < \, (H^{-1}M_P)^{{1\over 3}}\, M_P^{-1},
\end{equation}
we find from (\ref{nandm})  the corresponding maximal ${\bar n}$, by taking the integral up to 
$M_{BH} \, = \, (H^{-1}M_P)^{1/3} M_P$. Since, by default,  the  size of this BH is $\gg M^{-1}$, we have
 $MM_{BH} \, \gg \, M_P^2$ and the first term dominates in the last equation of (\ref{nandm}). This gives
 \begin{equation}
\label{nbig}
{\bar n} \, \sim \, {M_P \over M} (H^{-1}M_P)^{{1\over 3}}.
\end{equation} 
Again, ${\bar n}$ sets the maximal number of particles that can participate in the experiment, without 
making BH unacceptably long lived.  Now, if  $N \, > \, {\bar n}$, the following constraint emeges.  
Using a subset of ${\bar n}$ species and peforming the thought experiment with the BH formation and evaporation, we arrive to the usual flat space constraint 
\begin{equation}
\label{nbarbound}
{\bar n}  \, \lsim \, {M_p^2 \over M^2}. 
\end{equation}
From here, by taking into the account (\ref{nbig}), we get the following bound on the mass of the species
\begin{equation}
\label{massbound}
M \, \lsim \, {M_P \over (H^{-1}M_P)^{{1\over 3}}}.
\end{equation}

 \subsubsection{Constraint on Light Species:  $M \, \lsim \, H$} 
 
 In such a case,  even a BH as big as $\sim H^{-1}$, can have a lifetime $\sim H^{-1}$. So, for finding
 $\bar{n}$ in eq(\ref{nandm}) the integration must be  performed up to  
 $M_{BH} \, = \, H^{-1} M_P^2$.    Since ${M_P^2 \over M} \, \gg \, H \, \gg M$,  this gives, 
 \begin{equation}
\label{newnbar}
{\bar n} \, \sim  \, (H^{-1}M_P)^{2}.
\end{equation} 
Checking for the lifetime, we get 
\begin{equation}
\label{taubig}
\tau \, \lsim \, {1 \over {\bar n} } H^{-3}M_P^2 \, \sim \, H^{-1},
\end{equation}
which confirms the legitimacy of the derivation.  
 Then again, because  by default ${\bar n}$ has to satisfy the bound (\ref{nbarbound}), we get 
 \begin{equation}
\label{extra}
(H^{-1}M_P)^2 \, < \, M_P^2/M^2,
\end{equation}
 which is automatically compatible with the original assumption that $H \, > \, M$.  
 What remains is to be seen that  $N \leq {\bar n}$. This follows from the Gibbons-Hawking temperature 
 constraint. Indeed, because the de Sitter space  is a thermal bath with effective temperature  $T_{GH} \, \sim \,  H$,  the contribution to the energy density from   $N$ species with masses $M  \, < \, H$ would be 
\begin{equation}
\label{tspecies}
\rho_{species} \, \sim \, N \, H^4.    
\end{equation} 
 This contribution cannot exceed the energy density of the de Sitter vacuum,  which puts the upper bound on the number of species lighter than $H$, to be $M_P^2/H^2$.   Notice, that for species that are lighter than $H$, this is a more stringent bound, than the flat space one.

\subsection{ Classically Stable Vacua: Relaxing the Longevity Constraint}

 In the analysis of the previous section, we have deliberately limited ourselves by considering BH that are sufficiently short-lived.  This requirement is certainly justified for the time dependent vacua, which can only be regarded  as stationary de Sitter on the time-scales of few Hubble.   Most of the slow-roll 
 inflationary vacua fall in this category.  
  
 On the other hand,  the vacua  that correspond to the classically-stable minima of the landscape, are exponentially long lived.  For such vacua,  the requirement (\ref{lifetime}), demanding that the BH evaporation time to be less than the Hubble time,  is unnecessarily stringent.  Indeed, we can have a hypothetical observer orbiting around a 
 BH on a stationary orbit for much longer than the Hubble time.   What is important in such a case, is that 
 the lifetime of the BH is longer than the lifetime of the vacuum $\tau_{vac}$.  If latter is the case,  we
can relax the requirement 
 (\ref{lifetime}) and only demand (\ref{randh}). 
  It is again useful to split the discussion in two cases, corresponding to the mass of the species being heavier or lighter than the Hubble.
  
   \subsubsection{Constraint on Heavy Species:  $M \, \gg \, H$}
 
    Again we first have to find the number of available species ${\bar n}$, which can participate in the 
    BH formation and evaporation experiment  that are compatible with the constraint (\ref{randh}). 
    This can be found by integrating (\ref{nandm}) up to the mass of the Hubble size BH, which has a mass  $M_{BH} \, \sim \, H^{-1}M_P^2$.  
 This gives 
 \begin{equation}
\label{nheavy}
{\bar n} \, \sim \,  {M_P^2  \over  MH}. 
\end{equation}
 An alternative  way of finding the maximal number of heavy  species of mass $M \, \gg \, H$, that can participate in the experiment, is by estimating of  how many such particles can fit within the de Sitter horizon before turning  the Hubble volume into a BH, 
  \begin{equation}
\label{nbar1}
{{\bar n} M \over M_P^2} \, \sim \, H^{-1}\, ~~\rightarrow  \, ~~ {\bar n} \, \sim \, {M_P^2 \over HM}  \, .
\end{equation}
Because $M \, \gg \, H$, the above number is much larger than the flat space bound on the number of species. Thus, in this case, de Sitter is essentially not limiting the number of species that one could use in BH formation, and the flat space BH bound remains. Thus we have, 
\begin{equation}
\label{heavybound}
N \, \lsim \, {M_P^2\over M^2},
\end{equation}
just as in flat space. This makes perfect sense. Indeed,  in an eternal de Sitter space, sub-horizon BHs formed by the heavy particles, evaporate just as in the flat space. 
 
  \subsubsection{Constraint on Light Species:  $M \, \lsim \, H$}

 In this case  expression for ${\bar n}$ (again the maximal number of particles that can be used in experiment without conflicting with (\ref{randh})) changes to
\begin{equation}
\label{newnbar}
{\bar n} \, = \, (H^{-1}M_P)^{2}.
\end{equation} 
Let us find out, what is the constraint on $N$ in such a case. Let us first show, that we
cannot have $N\, > \, {\bar n}$ due to Gibbons-Hawking temperature argument. 
Because 
\begin{equation}
\label{mass}
M \, \lsim \, H,
\end{equation}
in the de Sitter space all the species contribute to Gibbons-Hawking radiation.    
 Each species with mass $< H$, 
will contribute into the thermal energy a factor $\sim \, H^{4}$, which for $N \, > \, {\bar n}$ would exceed the 
energy in de Sitter space. This is impossible.   Thus, we arrive to the conclusion that ${\bar n}$ 
is the bound on $N$. Thus,  
\begin{equation}
\label{Nbound}
N \, \lsim \, {M_P^2 \over H^2}\, .
\end{equation}
 Again, this result agrees with the general intuition, since in the presence of  sub-Hubble mass species, 
 the de Sitter horizon strongly limits the size of the BH that in the flat space 
 would evaporate into the light species. 
   Thus, the key point is that for the light species
$M \, \ll \, H$, the bound is cut-off by the Gibbons-Hawking temperature argument, which is more stringent than the flat space bound of \cite{dvali}.  

 We shall now apply this consideration to different inflationary scenarios. 
 
\section{Application for the Landscape} 

\subsection{Stationary SUSY-Breaking  de Sitter Vacua }

In this section, we shall apply our consideration to the vacua that are classically stable, and thus have an exponentially long life-time. 

  Consider a  nearly Minkowski vacuum in which  gravitino mass is $m_{3/2}$. In the standard picture 
  our MSSM vacuum is such.  In this vacuum there are moduli that are getting masses from the SUSY-breaking dynamics, and their masses are  $\sim m_{3/2}$.  These moduli parameterize the would be flat directions, that are lifted by SUSY-breaking.  When we move along the lifted flat directions, many particles become massive.  Let such modulus be $\phi$.  For example, $\phi$ can be one of the MSSM flat directions. In $F$-term type supersymmetry breaking,  the potential for moduli is generated through the K\"ahler couplings to the SUSY- breaking  $F$-terms and has a form
\begin{equation}
\label{vmod}
V(\phi) \, = \,  m_{3/2}^2 \, M_P^2  \, \mathcal{V} ({\phi \over M_P}).  
\end{equation}
 Usually, is it assumed that the function $\mathcal{V}(\phi/M_P)$ can have many new minima at values $\phi \sim M_P$.  However, 
the BH bound derived in the previous section can restrict such possibilities. 

 To see this, imagine that indeed there is a new minimum at $\phi \sim M_P$.  Of course, typically this minimum will not be Minkowski and will have a vacuum energy of order the SUSY-breaking scale 
  $V_0 \sim m_{3/2}^2M_P^2$.  The question is, what is the restriction on the number of species 
 of mass $M$  in such a vaccum. 
 
  In this section we will be interested in classically-stable vacua, which  can only decay through the tunneling  process and thus,  have an exponentially-long lifetime  $\tau_{vac}$.  We shall show, that 
 BH consideration  of the previous section can  provide a restriction on this lifetime in terms of number
 of species $N$ and their mass $M$.  For definiteness, we shall discuss vacua with $M \, \gg \, m_{3/2}$. 
 
   In order to see this, let us first  assume that the vacuum in question can be arbitrarily long lived.  In particular, $\tau_{vac}$ can be much longer than the lifetime 
 of a minimal BH satisfying the constraint (\ref{randh}).  That is 
   \begin{equation}
\label{tauvac}
\tau_{vac} \, \gg \, \tau_{BH}  \, \sim \,{(\bar n M)^3 \over M_P^4} \, \sim \,  {H^{-3}M_P^2},  
\end{equation} 
where in the last expression we have taken in to the account  (\ref{nheavy}). Then,  
 as shown in the previous section the  BH proof of the bound, $N M^2 < M_P^2$, will go through. 
The requirement that the gravitational radius $r_g$ of a minimal BH incorporating all the species, is less than the  curvature radius of the  vacuum ($H^{-1}$), applied to SUSY-breaking vacua, takes the form 
\begin{equation}
\label{bound}
r_g  \, \lsim \,   H^{-1} \,~~~  \rightarrow  ~~~~ \,  NM   \, \lsim \, M_P^2/m_{3/2},  
\end{equation}
where we have used the fact that the mass of a minimal BH (containing all the species) is $M_{BH} \, \sim \,  NM$.  Because 
(\ref{bound}) implies 
\begin{equation}
\label{bound1}
N  \, \lsim \,{ M_P^2 \over M^2} {M \over m_{3/2}} \, ,    
\end{equation}
the flat space BH bound
\begin{equation}
\label{bound2}
N  \, \lsim  \,{ M_P^2 \over M^2}   
\end{equation}
is automatically valid even in the curved vacua (with $V_0 \sim m_{3/2}^2M_P^2$),  as long as,  $M > m_{3/2}$.   

Now it is obvious that the above result  puts a severe restriction
on all the vacua, that are obtained my modular deformation from the Minkowski vacuum in which supersymmetry breaking scale is hierarchically small.  For instance, on the deformations of the standard 
MSSM vacuum in which the hierarchy problem is solved by the low energy SUSY-breaking. 
An immediate implication is that there cannot be the metastable vacua in which  MSSM flat directions
have $\gsim M_P$ VEVs, since  such vacua would automatically fall within the conditions of the BH proof, and in the same time there many species will get masses $M\sim M_P$, in contradiction with this 
bound.  The same is true for the deformations of the vacua with GUT symmetry breaking, and for many other cases.
 
  What happens if the bound is not satisfied, for example, what if there are too many massive particles? 
  Then, by consistency, theory has to respond by decreasing the  lifetime of the vacuum, in such a way that  (\ref{tauvac}) is no longer valid.  That is, a large number of species must destabilize the vacuum! 
In such a case the vacuum in consideration becomes short lived or even classically unstable, and 
the argument has to be reconsidered.  We shall discuss such a situation  in the next section. 
 
  \subsection{Constraint on the Slow-roll Inflationary States}. 
  
  The black hole bound on species (\ref{nmax}) can be extended not just to the (meta) stable
 vacua, but also to time dependent ``vacua",  with slowly changing values of the parameters. 
 The important examples from this class of vacua are the inflationary  slow-roll backgrounds. 
 We shall now  apply the BH bound to such states. 
 
  Consider a slow roll inflation driven by a single inflaton field $\phi$.  The equation for the spatially-homogeneous time-dependent field is, 
  \begin{equation}
\label{infeq}
\ddot{\phi}\, + \, 3H\, \dot{\phi} \, + \, V(\phi)'\, = \, 0\, ,   
\end{equation}
where, prime stands for the derivative with respect to $\phi$.      
   The main idea of the slow roll inflation is, that for certain values of  $\phi$, the potential $V(\phi)$ is sufficiently flat, so that the friction term dominates and this  allows $\phi$ to roll slowly.  
 The energy density is then dominated by the slowly-changing potential energy.  The 
 Hubble parameter is approximately given by $H^2 \, \simeq \, V(\phi) /3M_P^2$, and can be regarded as constant on the time scales $\sim \, H^{-1}$.   Obviously, the inflationary region of the potential must be away from  todays minimum with almost zero vacuum energy.  In any inflationary scenario the value of the inflaton field during inflation is very different from its todays expectation value $\phi_0$ corresponding 
 to the minimum of $V(\phi)$, which without loss of generality we can put at $\phi_0 \, = \, 0$.  
   
 Soon after the end of the inflationary period, inflaton oscillates about its true minimum $\phi_0$, and reheats 
 the Universe.   For this to happen,  inflaton should necessarily interact with the standard model particles and possibly with the other fields.  Let us consider an inflaton coupled to $N$ species, with masses 
 $M_j$.   For the efficient reheating, the masses of the the particles  
about the minimum $\phi_0$,  must be less than the inflaton mass about the same minimum. 
That is,  $M_j \, \ll \, V''(\phi_0)$.   Due to coupling to the inflaton field, the masses of species 
are functions of its expectation value,  $M_j(\phi)$, and it is very common that these masses change substanctialy during inflation.   The key point that we are willing to address now, is that the masses of these species are subject to the BH bound, and give useful restriction on the inflationary trajectory.  Thus, knowing the 
couplings of the inflaton in {\it our} vacuum,  one can get an non-trivial information about
the much remote inflationary vacua of the same theory. 

  For simplicity, we shall assume the universality of the species masses  $M_j(\phi) \, = \, M(\phi)$.  
 During the slow-roll inflation, Universe is in a quasi-de-Sitter state, in which the inflationary Hubble parameter sets the size of the causally-connected event horizon $H^{-1}$.   However, the difference 
 from the stationary de Sitter vacua, is that in realistic inflationary scenarios the slow roll phase 
 (in any given region)  is not exponentially long lived,  and lasts for several Hubble times.
 So $H^{-1}$ sets the time scale on which parameters can be regarded as constant. 
 
  Thus, a hypothetical observer  located within a given  causally-connected 
  inflationary patch can perform a sensible experiment with BH formation and evaporation, as long as 
  the gravitational radius $r_g$ and the BH lifetime $\tau$ obey the bounds (\ref{randh}) and (\ref{lifetime}).   In such a case, the considerations of section 1.1 can be directly applied, and we arrive to the bound, 
 \begin{equation}
\label{massbound}
M(\phi) \, < \, {M_P \over (H^{-1}(\phi)M_P)^{{1\over 3}}}.
\end{equation}
  All the information that this bound implies for a given inflationary scenario, is encoded in the functions 
$M(\phi)$  and  $H(\phi)$. We shall now illustrate this on some well known examples. 
 
  \subsection{Chaotic Inflation}
  
   Let us consider the example of Linde's chaotic inflation \cite{chaotic}. This is based on  a single scalar field 
   with a mass $m$ and no sef-coupling
   \begin{equation}
\label{chao}
V(\phi) \, = \, {1 \over 2} m^2 \phi^2 \, + \,  g \phi \bar{\psi}_j\psi_j\, .
\end{equation}  
The last term describes the coupling to $N$-species, which for definiteness we assume to be fermions, 
and $g$ is the interaction constant.   As said above, the coupling of the inflaton to the species is crucial
for the reheating. 

 The above theory has a Minkowski vacuum, in which $\phi \, = \, 0$ and  all the species are massless. 
 Due to the latter fact, in this vacuum  the BH bound on the  number and mass of the species is satisfied.  However, as we shall see,  the same bound, puts non-trivial restriction  on the inflationary epoch, since 
 during inflation  $\phi \,  \neq \, 0$ and species are massive.  

 Ignoring for a moment the coupling to the species, the logic  in the  standard Chaotic inflationary scenario goes  as follows.  The expectation value of the field $\phi$ can be arbitrarily large, as  long as the 
 energy density remains sub-Planckian, that is 
 \begin{equation}
\label{condition}
m^2\phi^2 \,  \ll \, M_P^4\, .  
\end{equation}   
 The equation (\ref{infeq}) then can be applied and takes the form 
  \begin{equation}
\label{cheq}
\ddot{\phi}\, + \, 3H\, \dot{\phi} \, + \, m^2\phi \, = \, 0\, ,  
\end{equation}
where $H^2 \, = \, {m^2\phi^2 \, + \, \dot{\phi}^2 \over 6M_P^2}$. As long as $H \, \gg \,  m$, the friction dominates and $\phi$ rolls slowly.  This implies (up to a factor of order one)  
\begin{equation}
\label{slow}
\phi \,  \gg \, M_P,
\end{equation}
which is compatible with (\ref{condition})  as long as $m \, \ll \,  M_P$. 
If the above is satisfied, $\phi$ rolls slowly, and Universe undergoes the exponentially fast expansion. 
Let us now see how the coupling to the species restricts the above dynamics. During inflation the mass of the species is $M \, = \, g\phi$ and they are subject to the BH bound.  To see what this bound implies we can simply insert the current values of $M(\phi)$ and $V(\phi)$  in  (\ref{massbound}), and we get  
 \begin{equation}
\label{massbound1}
g\phi \, \lsim \, M_P\left ({m\phi  \over M_P^2}\right )^{{1\over 3}}.
\end{equation}
Non-triviality of the above constraint is obvious. For example, the standard argument assumes  that 
inflation could take place for arbitrary $m \, \ll \, M_P$, and from arbitrarily large values of $\phi$ satisfying
(\ref{condition}), irrespective to the number of species to which inflaton is coupled.
The above expression tells us that in the presence of species,  this is only possible, provided, 
$g \, \lsim \, (M_P/\phi)^{2/3} (m/M_P)^{1/3}$. 

For the practical reasons of solving  the flatness and the horizon problems, in the standard Chaotic scenario, last $60$ e-foldings happen for $\phi \,  \lsim \,  10 M_P$, whereas 
from density perturbation we have  $m \, \sim  10^{12}$GeV or so.   This implies, 
$g \, < \, 10^{-3}$. This constraint can be easily accommodated by the adjustment of couplings, however  it is remarkable that no fine tuning can make $g \sim 1$ consistent.

 \subsection{Hybrid Inflationary Vacua} 
 
 The essence of the  hybrid inflation \cite{hybrid}  is that inflationary energy density is not dominated by the potential 
 of the slowly-rolling inflaton field $\phi$, but rather by a false vacuum energy of other scalar fields, $\chi_j$. These fields are  trapped in a temporary minimum, created 
due to large positive mass$^2$-s, which  they acquire  from the coupling to the inflaton field.  The slowly rolling inflaton then acts as a clock,  which at some critical point triggers the transition that liberates the trapped fields, and converts their false vacuum energy into radiation.  However, usually  Inflation ends before this transition, because 
of breakdown of the slow-roll.  Thus, in hybrid inflation, the presence of fields with inflaton-dependent masses is essential not only for the  reheating, but for the inflation itself.  

The simplest prototype model 
realizing this idea is  
  \begin{equation}
\label{hybrid}
V\, =\, \lambda^2 \, \phi^2 \chi_j^2 \,  + \, \left ({g \over 2} \,  \chi_j^2 \, - \, \mu^2 \right )^2,
\end{equation}
where $\lambda$ and $g$ are constants. 
Then,  for $|\phi| \, > \, \phi_t\, \equiv \, \mu \sqrt{{g \over \lambda^2}}$,  the effective potential for 
$\chi_j$ is minimized at $\chi_j = 0$, and the false vacuum energy density is a $\phi$-independent  constant, $\mu^4$.  Thus, in the classical treatment of the problem, starting at arbitrary initial value $\phi \, \gg \, \phi_t$ and with zero initial velocity, $\phi$ would experience zero driving force and system would inflate forever.  One could slightly lift this flat direction by adding an appropriate self interaction potential 
for $\phi$ (e.g., such as a positive mass term $m^2\phi^2$) which would drive $\phi$ towards the small values.  In such a picture inflation ends abruptly after $\phi$ drops to its critical value $\phi_t$, for which $\chi_j$ becomes tachionic, and system rapidly relaxes into the true vacuum.  
However, the above story is only true classically, and quantum mechanical corrections 
are very important and always generate potential for $\phi$ \cite{fterm2, dterm1}. Because of to these corrections, typically, inflation ends way  before the phase transition, due to breakdown of the slow-roll.  Existence of supersymmetry cannot change the latter fact, however, supersymmetry  does make
the corrections to the potential finite and predictive. 

 The simple supersymmetric realizations of the hybrid inflation idea have been suggested in form of $F$-term \cite{fterm1,fterm2}
 and  $D$-term \cite{dterm1,dterm2} inflationary models.   As a result of supersymmetry, in $F$-term inflation $\lambda\, = \, g$. As it was shown in \cite{fterm2} and \cite{dterm1}, due to renormalization 
of the K\"ahler function via $\chi_j$ loops, the non-trivial inflaton potential is inevitably generated, which for $\phi \gg \phi_t$ 
has the following form, 
\begin{equation}
\label{oneloop}
V(\phi) \, \simeq \, \mu^4 \left [ 1 \, + \, {N g^2\over 16 \pi^2} {\rm ln}{g|\phi| \over Q}  \right ] \,, 
\end{equation}
where, $Q$ is the renormalization scale.  Notice, that  this potential cannot be fine tuned away by addition of some local counter terms.  The condition of the slow roll is that  $V'' \ll H^2$, implying that  
\begin{equation}
\label{slowroll}
N\, g^2 \, \ll \, {\phi^2 \over M_P^2}\, . 
\end{equation}
Because of the logarithmic nature, the slope flattens out for large $\phi$. 
However, even if one tries to ignore any other correction to the potential,  
nevertheless, the slow-roll condition will eventually run in conflict with the black hole bound,
which implies that 
\begin{equation}
\label{conflict}
N\, g^2 \, \lsim \,  {M_P^2 \over \phi^2} \,.  
\end{equation} 
This fact indicates, that even if the theory is in seemingly-valid perturbative regime (that is, 
$  {N g^2\over 16 \pi^2} {\rm ln}{g\phi \over Q}  \, \ll  \, 1$), nevertheless, the perturbative corrections 
to the K\"ahler cannot be the whole story, and theory has to prevent growth of $\phi$, by consistency with 
the black hole physics. 
 
  We wish to point out one subtle difference between the $F$-term and $D$-term inflationary scenarios. 
 In case of $F$-term inflation, $\chi_j$ fields need not transform under any long range (un-Higgsed) 
 gauge symmetry.  However, in case of the $D$-term inflation story is more involved, because the 
 mass parameter $\mu^2$  comes from the Fayet-Illiopoulos term $\xi $ of an $U(1)$ vector supermultiplet.  In the globally supersymmetric limit, the potential has the form \cite{dterm1} 
 \begin{equation}
\label{hybridDterm}
V\, =\, \lambda^2 \, |\phi|^2 \left (|\chi_j|^2 \, + \, |\bar{\chi}_j|^2 \right )\,  + \, {g^2\over 2} \left ( \,  |\chi_j|^2 \, - \, |\bar{\chi_j}|^2 \, - \, \xi^2 \right )^2,
\end{equation}
where, $\chi$ and $\bar{\chi}$ carry opposite charges, which we take equal to $+1$ and $-1$ respectively.  The mass of the $U(1)$ gauge field (call it $W_{\mu}$)  therefore vanishes above the critical point
$|\phi|^2 \, > \, \phi_t^2 \, \equiv \,{g^2 \over \lambda^2}  \xi$.  However, this is an artifact of the global supersymmetry. 

 The most important effect of supergravity corrections to this picture is that  
 $U(1)$ becomes a gauged $R$-symmetry \cite{gauger}, and the charges experience a shift of order $\xi/M_p^2$.  
 This can be seen from the expression for the
covariant derivative on the gravitino ( we use conventions of \cite{fiterm}, see details there)
 \begin{equation}
 {\cal D}_{[\mu }\psi _{\nu ]}=\left(
\partial _{[\mu} +{1\over 4} \omega _{[\mu} {}^{ab}(e)\gamma _{ab} +
{1\over 2}{\rm A}_{[\mu} \gamma _5\right)\psi _{\nu ]}\,,
 \label{covdergr}
\end{equation}
where $\omega _{\mu}^{ab}(e)$ is the spin connection, and  the $U(1)$-connection $A_\mu$ is given by
\begin{equation}
  A_\mu =\frac{1}{2}
  \left[ (\partial _i{\cal K})\hat \partial _\mu z^i-(\partial
  ^i{\cal K})\hat \partial _\mu z_i\right]
  +\frac{g\xi}{M_P^2}W_\mu  \,,
  \label{AmuB}
\end{equation}

where
\begin{eqnarray}
\hat \partial _\mu z_i=  \partial _\mu z_i- W_\mu\, \eta_i(z)\,.
\end{eqnarray}
Here, ${\cal K}$ is the K{\"a}hler function, and sum runs over all the chiral
superfields $z_i$ and $\eta_i(z) $ are the holomorphic functions that set
the  $U(1)$  transformations of  all chiral superfields in the
superconformal action,
\begin{equation}
  \delta  z_i= \eta_{i}(z)\alpha(x)\, .
\label{eta}
\end{equation}
When the K{\"a}hler potential is $U(1)$-invariant, as is the case in the simple model above, 
 the $U(1)$ gauge
transformation of the gravitino gauge-connection $A_\mu $ takes a
universal form:
\begin{equation}
 \delta  A_\mu = \frac{g\xi}{M_P^2} \delta  W_\mu = \frac{g\xi}{M_P^2} \partial_\mu \alpha(x) \, , 
  \label{deltaconnection}
\end{equation}
which means that
gravitino acquires an  $U(1)$-charge,  and thus $U(1)$ becomes  an
$R$-symmetry.

 Because of this charge shift, it is not at all guaranteed that $U(1)$ will stay un-Higgsed even though 
 $\chi_j$ VEVs vanish. The characteristic mass of the $U(1)$ photon is at least as large as the Hubble
 parameter.  
    As we shall discuss,  this is exactly what happens in $D$-brane inflation.

 \subsection{Brane Inflationary Vacua} 
 
  A possible mechanism for the inflation in string theory, is brane inflation 
  \cite{brane1,brane2,brane3,brane4,Kachru:2003sx,Hertzberg:2007wc}. 
  In this picture the role of the inflaton field $\phi$ is played by the brane-separation field. A simplifying but crucial assumption of the original brane inflation model, is that compactification moduli are all fixed, with the masses being at least of order of the inflationary Hubble parameter, so that 
branes can be considered to be moving in a fixed external geometry, weakly affected by the brane motion.  In the same time, the  $4d$ Hubble volume must be larger than the size of the compact extra dimensions. These conditions allow us to apply 
 the  power of the effective four-dimensional supergravity reasoning. 
 
  Below we shall focus on the case of $D-brane$ inflation, based on the motion and subsequent annihilation of branes an anti-branes.  In \cite{brane3},  it was shown that  this picture 
 from the four-dimensional perspective can be understood as the hybrid inflation, in which
 $\phi$ is a brane distance field, and role of $\chi$ is played by the open string tachion.  
 
   An interesting evidence, indicating that  $D$-brane inflation as seen from the $4d$ supergravity 
 perspective is of the $D$-term type, emerged later (see \cite{dd,fiterm}).  This connection allows us to apply the above-derived black hole constraints to brane inflation both from $4d$ supegravity as well as 
 from  $10d$ string theory point of view. 


 In this picture, the supersymmetry breaking by a non BPS
brane-anti-brane system  corresponds to the spontaneous supersymmetry
breaking via FI $D$-term. 


When branes are far apart, there  is a light  field $\phi$, corresponding
 to their relative motion. This mode is a
combination of the lowest lying scalar modes of the open strings that are
attached to a brane or anti-brane only. 
We are
interested in the combination that corresponds to the relative radial
motion of branes.

\begin{equation}
\label{X} \phi \, = \, M_s^2 r\,,
\end{equation}
where $M_s$ is the string scale.

 In the simplest case of a single brane-anti-brane pair,  we have the two gauged $U(1)$-symmetries.
 One of these two  provides a non-vanishing
$D$-term.  The tachyon ($\chi$) is an open string state
that connects the brane and the anti-brane.  The mass of this stretched
open string is $M_s^2 r$. In $4d$ language,  the tachyon as
well as other open string states get mass from the coupling to $\phi$.

 The energy of the system is given by the
$D$-term energy, which is constant at the tree-level, but not at one-loop
level. At one-loop level the gauge coupling depends on $\phi$.  $g^2$ gets renormalized,
because of the loops of the heavy $U(1)$-charged states, with $\phi$-dependent masses.  
 For instance,
there are one-loop contributions  from the $\chi$ and $\bar\chi$ loops.
More precisely there is a renormalization of $g^2$ due to one-loop open
string diagram, which are stretched between the brane and anti-brane.
Since the mass of these strings depend on $\phi$, so does the renormalized
$D$-term energy
\begin{equation}
\label{DX} V_D \, = \, {g^2(\phi)\over 2}D^2  \, = \, {g^2_0 \over 2} \left ( 1 \,
+ \, g^2_0f(\phi)\right ) \xi^2\,,
\end{equation}
where $g_0^2$ is the tree-level gauge coupling, and $f(\phi)$ is the
renormalization function.  For example, for $D_3-D_7$ system \cite{d3d7} at the intermediate distances 
($M_s^{-1} \ll r \ll R$, where $R$ is the size of two transverse extra dimensions), 
this takes the form (\ref{oneloop}). 


 We shall now see, why  at  least in the simplest $D$-brane setup, the $U(1)$ symmetry must be 
 Higgsed throughout the inflation. 

Let us again think about the  process of  D$_{3
+ q}-\bar{\mbox{D}}_{3+q}$ driven inflation, with the subsequent brane annihilation.  We assume that $q$
dimensions are wrapped on a compact cycle, and relative motion takes place in $6 - q$ remaining
transverse dimensions. 

The low energy gauge symmetry group is $U(1)\times U(1)$, one linear
superposition of which is Higgsed by the tachyon VEV.
The crucial point is, that this Higgsed $U(1)$
gauge field is precisely the combination of the original $U(1)$-s that
carries a non-zero RR-charge (the other combination is neutral).  The
corresponding gauge field strength ($F_{(2)}$) has a coupling to the closed string RR
 $2+q$-form ($C_{(2\, + \, q)}$) via the WZ terms,
\begin{equation}
\label{fccoupling} \int_{3 + 1 + q} F_{(2)}\wedge C_{(2 \, +\, 
q)}\,,
\end{equation}
where, since we are interested in the effective $4d$ supergravity
description,  we have to integrate over extra $q$-coordinates, and only keep the $4d$ zero mode component of the RR field. This then becomes an effective $2$-form, $C_{(2)}$. 

The connection with the $4d$ supergravity  $D$-term  language,
is made by  a
dual description of the $C_{(2)}$-form in terms of an axion ($a$),
\begin{equation}
\label{cduala}
d {\rm C}_{(2)} \rightarrow *\, d {\rm a}\,,
\end{equation}
where star denotes a $4d$ Hodge-dual. Under this duality transformation
we have to replace
\begin{equation}
\label{c-a} 
(d {\rm C}_{(2)})^2  \, + \, {\xi \over M_P^2} F_{(2)}\wedge {\rm C}_{(2)}\, ~~
 \rightarrow \,~~ M_P^2 (d{ \rm a \, - \,g Q_a
W})^2\,,
\end{equation}
where $Q_a \, = \, {\xi \over M_P^2} $ is the axion charge under $U(1)$.
As it should, this charge vanishes as the compactification volume goes to infinity, and
$4d$ supergravity approaches the rigid limit\footnote{The above value of the
axionic charge, reproduces the correct RR charge
of the $D_1$-string, and also has a correct scaling for the anomaly
cancellation\cite{fiterm}.}.
  We thus see that the $U(1)$ gauge field ($W_{\mu}$) acquires a mass $m_{W}^2 \, \gsim \xi^2 /M_P^2$.  

 We are now ready to discuss applicability of our BH thought experiment to the above 
 $D$-brane system.   Since the role of the species $\chi_j$, that are getting mass from the inflaton field, 
 is played by the stretched open strings, the first condition for the applicability of the BH bound is, that  these strings should fit at least  within the Hubble size black hole. This is automatically the case, 
 since by the validity of the brane inflation, the effective $4d$ Hubble volume must be much larger than the size of  the compactified dimensions.   Since the length of the stretched strings cannot exceed the latter size, they automatically fit within the black hole  horizon.
  
  The second issue is the possible interference of the $U(1)$ ``hair" of the open string tachion with the black hole  formation and evaporation process.  Again, as we have seen, the black holes of interest have size of order Hubble, 
 which is comparable to the Compton wavelength of the $U(1)$-photon. On the other hand, stretched 
 strings are heavy, so  the lifetime 
 of such a black hole is many Hubble times.  Typical time scale for a black hole to loose a 
 photon hair  is the Hubble time (because of photon mass), after this time, black holes should evaporate 
 as normal hairless black holes.  So again, at least to leading order, the massive $U(1)$ 
 photon should not interfere with our arguments. 
 
  We should stress, however, that because the photon mass is roughly the same order as the curvature scale, more careful analysis would be very useful. This will not be attempted  here.

\vskip0.5cm


\noindent
{\bf Acknowledgments:}

\noindent
  We would like to thank  C. Kounnas, A. Linde, H. Nielsen, M. Petropoulos, M. Redi, and D. Tsimpis
  for stimulating discussions.   This work was supported in part by
  David and Lucile  Packard Foundation Fellowship for  Science and Engineering, by NSF grant PHY-0245068,  by  the EU under the contracts MEXT-CT-2003-50966 and by the Excellence Cluster "The Origin and the Structure of the Universe"
  in Munich.
  
\noindent

{\bf Note Added:}

 Before submitting this paper, ref. \cite{huang} appeared, which discusses some constraints on
$\phi^4$-inflation in the presence of species from a different perspective.

\bibliographystyle{plain}

\end{document}